\documentclass[useAMS,usenatbib]{emulateapj} \usepackage{epsfig}
\usepackage{lineno}
\usepackage{epsfig}  \usepackage{amsmath} \usepackage{amssymb}
\usepackage{natbib} \usepackage{graphicx} \usepackage{endnotes}
\usepackage{color} 

\shorttitle{AstroSat/LAXPC observation of Cygnus X-1} \shortauthors{ Misra et
al.}

\begin{document}

\title{ {\it AstroSat}/LAXPC observation of Cygnus X-1 in the hard state} 
 \author{ Ranjeev
Misra$^1$,J S Yadav$^2$, Jai Verdhan Chauhan$^2$, P C Agrawal$^3$, H M Antia$^2$,
Mayukh Pahari$^1$, V R Chitnis$^2$, Dhiraj Dedhia$^2$, Tilak Katoch$^2$,
P. Madhwani$^2$, R K Manchanda$^4$, B Paul$^5$, Parag Shah$^2$} \affil{$^1$ Inter-University Centre for
Astronomy and Astrophysics, Pune 411007, India
\texttt{rmisra@iucaa.in}} \affil{$^2$ Tata Institute of Fundamental
Research, Homi Bhabha Road, Mumbai, India} \affil{$^3$ UM-DAE
Center of Excellence for Basic Sciences, University of Mumbai, Kalina,
Mumbai-400098, India}  \affil{$^4$ University of Mumbai, Kalina,
Mumbai-400098, India} \affil{$^5$ Dept. of Astronomy \& Astrophysics,
Raman Research Institute,  Bengaluru-560080 India }

\centerline{\bf Accepted for publication in Astrophysical Journal; 27/12/16}

\begin{abstract}
We report the first analysis of data from {\it AstroSat}/LAXPC observations
of Cygnus X-1 in January 2016. LAXPC spectra reveals that the
source was in the canonical hard state, represented by a prominent
thermal Comptonization component having a photon index of $\sim 1.8$
and high temperature $kT_e > 60$ keV along with weak reflection and { possible} disk
emission. The power spectrum can be characterized by two broad lorentzian
functions centered at $\sim 0.4$ and $\sim 3$ Hz.  The r.m.s of the
low frequency component decreases from $\sim 15$\% at around 4 keV to 
$\sim 10$\%
at around 50 keV, while that of the high frequency one varies less
rapidly from $\sim 13.5$\% to $\sim 11.5$\% in the same energy range.
The time lag between the hard (20--40 keV) and soft (5--10 keV) bands varies
in a step-like manner being nearly constant at $\sim 50$ milli-seconds 
from $0.3$ to $0.9$ Hz, decreasing to   $\sim 8$ milli-seconds from
 $2$ to $5$ Hz and finally dropping to  $\sim 2$ milli-seconds for
higher frequencies. The time lags increase with energy for both the
low and high frequency components. 
The event mode LAXPC data allows for flux resolved 
spectral analysis on a time-scale of 1 second, which clearly shows that the
photon index increased from  $\sim 1.72$ to $\sim 1.80$ as the flux increased
by nearly a factor of two. We discuss the results in the framework of the fluctuation
propagation model.  
\end{abstract}


\section{Introduction}\label{intro}

Over the last several decades the persistent black hole system, Cygnus X-1,
has been a primary target of nearly all the X-ray observatories.
Its brightness and persistent nature makes it the ideal choice to understand
the behavior of black hole systems. 
Indeed, some of the earliest  paradigms were based on observations of
Cygnus X-1 such as the concept of an hot inner flow
where the thermal plasma  Comptonizes photons from a truncated standard 
disk producing
the characteristic hard spectrum \citep{Sha76} and even the alternative that 
the spectrum is produced by a corona on top of an accretion disk \citep{Lia77}.
The various black hole system spectral
states that are now known \citep[e.g.][]{Rem06} can be said to be 
inspired by the early classification of the spectra of 
Cygnus X-1 into hard and soft states \citep[e.g.][]{Tho75}.

{  Although Cygnus X-1 is a persistent high mass X-ray binary, it shares
several common spectral and timing features with other 
low mass X-ray binaries that harbor black holes.
For example, its hard state spectrum can be represented approximately by being
dominated by an hard power-law with a high energy cutoff, with weak reflection
and disk components which is also true for other black hole systems. As 
mentioned above, such a  hard state spectrum is consistent with the picture of a truncated standard accretion disk with an hot inner region giving rise to
the standard paradigm regarding the hard state geometry \citep{Sha76}. However
with the availability of high resolution spectra from {\it Suzaku} and 
{\it Nustar} this paradigm is now being challenged. Apart from absorption effects due to the ionized wind, the hard state spectra of
Cygnus X-1 show a relativistically broadened Iron line which indicates that
the standard disk is not truncated but extends all the way to the last stable
orbit \citep{Par15}. Such broad lines have also been reported for the luminous hard state of another black hole system GX 339-4 \citep{Pla15} suggesting that the geometry of the system may depend on the luminosity of the hard state.
Along with such detailed spectral studies, it is important to consider 
the timing properties of these sources to pin down the true geometry and
nature of these sources. }

The Rossi X-ray Timing Experiment (RXTE) has revolutionized our
understanding of the high frequency variability of black hole systems.
It is not surprising that one of the pioneering works, where the application
of timing analysis such as coherence, energy and frequency dependent 
time-lags were introduced was  an analysis of early RXTE data of
Cygnus X-1 \citep{Now99}. Since then a large number of analysis
has been undertaken on several RXTE data sets of Cygnus X-1 and now
results of  comprehensive timing analysis using nearly all 
RXTE observations of the source are available \citep[e.g.][]{Wil04}.

RXTE analysis have shown that for the hard state 
the variability of Cygnus X-1  is primarily between 
$\sim 0.1$ and $\sim 10$ Hz and  its power spectrum
can be approximately described by two to four broad lorentzians
in this frequency range \citep[e.g.][]{Now99,Wil04}.
 Time lag of high energy
photons with respect to the low energy ones are known to approximately
 increase linearly  with logarithm of the energy. The lags decrease 
as a function of frequency roughly as a power-law. Cygnus X-1 is known
to show a linear relation between flux and r.m.s  \citep{Utt01,Gle04}
 which is now known as the universal
flux-rms relation applicable to several X-ray binaries and 
Active Galactic Nuclei. Another related nearly universal property of
these systems, that the flux distribution is a log-normal one,
 was first shown for Cygnus X-1 \citep{Utt05}.

The time-scales of the variability as well as those of the time
lags are significantly longer than the light crossing time of the
inner regions of the accretion disk. This indicates that the variability
is related to the viscous time-scale suggesting an origin in the
outer regions of the disk. Yet, the variability is observed in hard
X-rays which originate from the inner regions. An explanation
was provided by \citet{Lyu97} where the accretion disk undergoes stochastic
variations all across the disk and
the fluctuations propagate inwards reaching the inner disk after a viscous
time-scale corresponding to their origin. Since the different fluctuations
have a multiplicative effect on the inner regions, the model can naturally
explain the flux-rms relation as well as the log normal distribution of
the flux. Moreover, the time lags as a function of 
energy and frequency have natural explanation in this framework
\citep[e.g.][]{Bot99,Mis00,Kot01}.
These successes have warranted
the construction of a detailed model (``PROPFLUC'')
 by which the predictions can
be quantitatively compared with observations especially the shape
of the power spectra at different energies as well as the time-lags
as function of energy and frequency \citep{Ing11,Ing12,Ing13,Rap16}.

The Large Area X-ray Proportional Counter (LAXPC) \citep{Yad16,Agr16} on
board the multi-wavelength satellite {\it AstroSat} \citep{Agr06,Sin14} has
several advantages over the Proportional Counter Array (PCA) of RXTE.
The effective area of LAXPC above 30 keV is significantly higher
than the PCA. The event mode data obtained from LAXPC allows
for studying the variations in user defined energy bins and for
flux resolved spectroscopy. Finally, the various instruments on
board {\it AstroSat} can provide wide band spectral coverage. LAXPC data
of another black hole system GRS1915+105 has already revealed the
capabilities of LAXPC to study  high frequency variability of
high energy photons \citep{Yad16b}.  

Here, we report on the first analysis of LAXPC observations of Cygnus X-1
to highlight its potential to make a significant enhancement in our
understanding of the system.

\section{LAXPC observations of Cygnus X-1}

\begin{figure}
\centering
\includegraphics[width=0.34\textwidth,angle=-90]{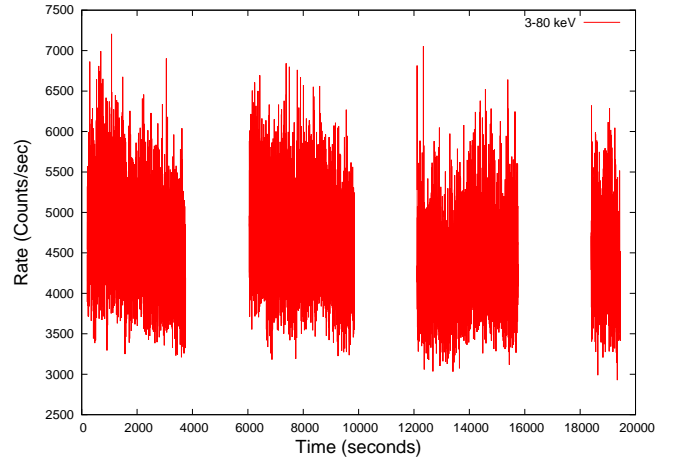}
\caption{Lightcurve of Cygnus X-1 in the energy
range 3--80 keV is shown where the count rate from all three LAXPC
detectors --- {\tt LAXPC10} , {\tt LAXPC20}  and {\tt LAXPC30} 
 are
combined. The gaps are due to SAA or earth occultations.}
\label{lightcurve}
\end{figure}

Figure \ref{lightcurve} shows the 1 second binned lightcurve generated using
data from three {\it AstroSat} orbits during Jan 8 2016. Cygnus X-1 is detected 
at a level of $\sim 5000$ c/s, except when the satellite passes through the
South Atlantic Anomaly (SAA) and when the source is obscured by the earth.
Data was used when the source was visible leading to an effective exposure
time of 12 ksecs.

\begin{figure}
\centering
\includegraphics[width=0.34\textwidth,angle=-90]{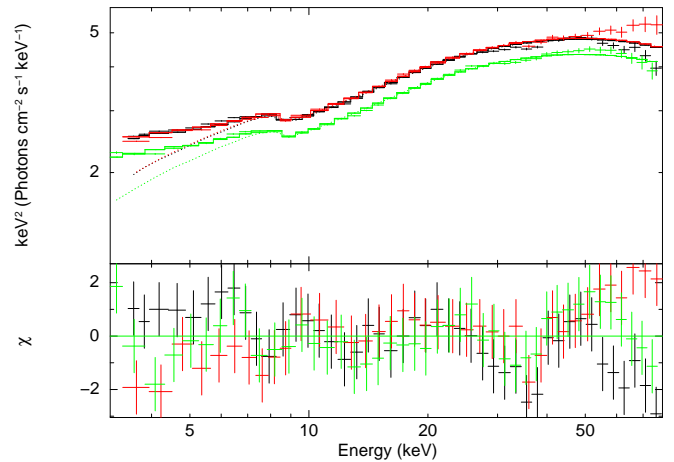}
\caption{The  3--80 keV unfolded spectra and residuals for the  three LAXPC
detectors --- {\tt LAXPC10} (red), {\tt LAXPC20} (green) and {\tt LAXPC30} (black) fitted jointly
by the model whose parameters are listed in Table \ref{specpar}.}
\label{spec}
\end{figure}

\begin{table}
 \centering
 \caption{Spectral Parameters}
\begin{center}
\begin{tabular}{ll}
\hline
  Model parameters &   Values \\ 
\hline
$\Gamma$ &  $1.77_{-0.007}^{+0.006}$  \\ 
&\\
$kT_{e} $ (keV) & $70^f$   \\ 
&\\
$kT_{disk}$ (keV) & $0.85_{-0.09}^{+0.09}$  \\
&\\ 
$N_{disk}$ & $235_{-74}^{+153}$  \\ 
&\\
$R_{refl}$ & $0.41_{-0.03}^{+0.04}$  \\
&\\
$\xi$ & $96_{-22}^{+34}$\\
&\\
Flux$_{3-80\; keV}$ ($\times 10^{-8}$ ergs/s/cm$^2$) & $1.751_{-0.007}^{+0.007}$\\
&\\
$\chi^2$/d.o.f & 134.2 / 120 \\
\hline
\end{tabular}
\tablecomments{$\Gamma$ and $kT_{e}$ are the photon index and electron temperature of the thermal Comptonization component;$^f$  The temperature, $kT_{e}$ was not constrained by the data and only a lower limit of 60 keV was attained. Hence
the temperature was fixed at
70 keV; $kT_{disk}$ and $N_{disk}$ are the
normalization of the disk component;  $R_{refl}$ and $\xi$ are the reflected fraction
and ionization parameter of the reflected component and Flux$_{3-80 keV}$ is the
unabsorbed flux in the 3-80 keV band. All errors are at the 2-sigma level.   }
\end{center}
\label{specpar}
\end{table}

The spectra were extracted from each proportional counter of the LAXPC
which are named as LAXPC 10, 20 and 30. The channels were grouped so that
the energy bins were about 6\% of the mean energy, which is roughly a third
of the average spectral resolution of the detectors i.e. $\sim 18$\%. The
spectra were fitted using XSPEC version 12.8.1. The energy range considered
was between 3 to 80 keV and all layers were included. The response and
background files were obtained from software that would  become part of the
LAXPC software release \citep{Ant16}. The background estimates were based on observed blank 
skies and at present, the background estimate is approximate.
Hence, its uncertainty was taken into account by adding 
a 5\% systematic error to the error on the background rate. 
The uncertainties in the
spectral response was taken into consideration by adding a systematic of 2\%
for the over all fitting. For the joint fitting of the three
proportional counters the relative normalization of LAXPC 20 and 30 were 
allowed to vary with respect to that of LAXPC 10. For the joint fit described
below,  the relative normalizations turned out
to be 1.10 and 1.11 of LAXPC 30 and 20 with respect to LAXPC 10.

The typical hard state spectrum of Cygnus X-1 in the 3-80 keV is 
known to be described
by a thermal Comptonization component with photon index $\Gamma \sim 1.8$,
a moderate reflection component and a weak disk emission. These
spectral components were described by the thermal Comptonization
model ``nthcomp'', the disk black body model ``diskbb'' and the
convolution reflection model ``ireflect'' in XSPEC. The seed photons
for the thermal Comptonization was assumed to be from the disk. The temperature
of the reflecting medium was taken to be the maximum temperature of the disk,
while its abundance was fixed to solar values. The fit was found to be insensitive to
the inclination angle and hence its cosine was fixed at 0.5. The spectrum above
3 keV is not sensitive to Galactic absorption and its column density was
fixed to $7 \times 10^{21}$ cm$^{-2}$ {  \citep[e.g.][]{Tom14}}
using the XSPEC model ``Tbabs''.
The best fit parameters obtained by the joint fitting are listed in Table
\ref{specpar} and the unfolded spectra with 
residuals are shown in Figure \ref{spec}.  {  The photon index of the
primary component is $\sim 1.77$. The upper limit of the 
electron temperature was not constrained while its lower limit was 60 keV. 
Hence we fixed the temperature to 70 keV. A weak disk emission is 
technically required by the data with a flux contribution of $\sim 15$\% 
in the 3-5 keV band. Its temperature is $\sim 0.85$ keV which is significantly
larger than $\sim 0.1$ keV obtained using {\it Suzaku} and {\it Nustar} data
\citep[Table 4 of][]{Par15}. More importantly its normalization of $\sim 200$
implies a very small inner radius of few kms for a distance of 2 kpc. This
discrepancy is not unexpected since the $4-10$ keV band is known to be complex
with ionized absorption and relativistically blurred Iron line \citep{Par15}.
 Clearly, such detailed spectral analysis is neither warranted nor possible 
for an instrument with LAXPC spectral resolution especially given that in these
early stages, the response and background are fairly uncertain. Our motivation
here is to show that  at this
level of uncertainties in the response and background, the data is well
represented by a typical hard state spectral model of Cygnus X-1}. Thus, 
the energy dependent  variability which is the primary focus of this work
can be considered to be reliable.

\begin{figure}
\centering
\includegraphics[width=0.34\textwidth,angle=-90]{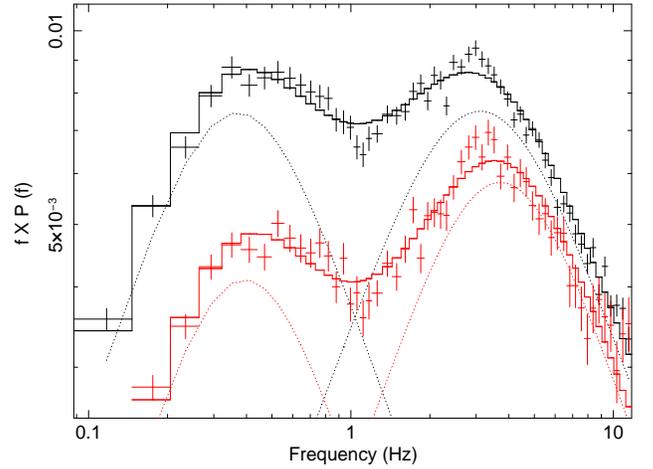}
\caption{The frequency times the power spectra of Cygnus X-1 for photons in the 3-10 keV band
(black points) and for those in the 20-40 keV band (red points). The spectra
has been fitted by two lorentzian components and the Figure shows the
changes in the power spectral shape as a function of energy. }
\label{powspec}
\end{figure}

\begin{figure}
\centering
\includegraphics[width=0.34\textwidth,angle=-90]{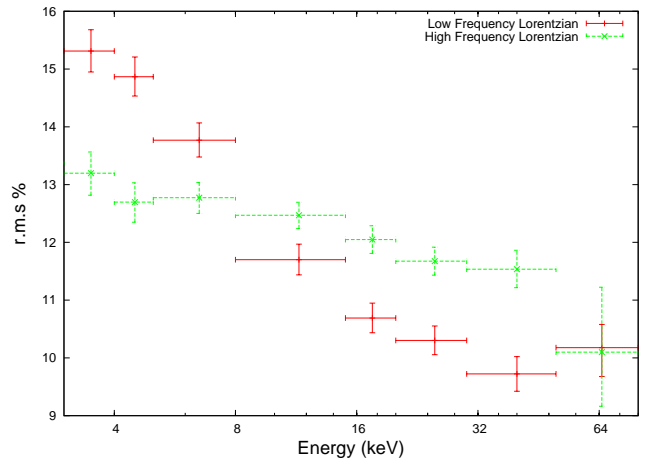}
\caption{The r.m.s of the low (red) and high (green) frequency
components as a function of energy. While the strength of the
low frequency components drops rapidly with energy that of the
high frequency is relatively more constant.  }
\label{rmsene}
\end{figure}

We compute the power spectra for different energy bands and show
the result in Figure \ref{powspec} for the 3-10  and 20-40 keV bands.
{  The lightcurve was divided into 702 segments each having a 
length of 1024 points with a time bin of 16.67 milli-secs.
The power spectrum was computed for each segment and then averaged. The
resultant power spectrum was rebinned in frequency space.}  The
dead time corrected Poisson noise assuming a dead time of 42 microseconds
\citep{Yad16b} has been subtracted and the power spectra have been
corrected for the corresponding background rates. The power spectra clearly show
two broad features at $\sim 0.4$ and  $\sim 3$ Hz and we use two
broad lorentzians {  \citep[e.g.][]{Bel02}} to empirically represent them. For the low and high energy
band the fit gives $\chi^2/dof$ of $130.2/52$ and $92.1/51$ respectively. 
If a systematic of 5\% is added, these $\chi^2$ values decrease to 
$39.4$ and $52.6$. Alternatively, one can instead, include two extra lorentzian
components and such a fitting gives acceptable reduced $\chi^2$ close to
unity. However, in the absence of a detailed physical model for the
components, in this work, we consider only the fit using  two Lorentzian and refer to them as the low and high frequency components.
For the low energy band
the low frequency is modeled as a lorentzian with centroid frequency
$0.17\pm 0.04$ Hz and  width $0.65\pm 0.06$ Hz. While
for the high energy band these values are found to be  $0.20\pm 0.05$ Hz and $0.66\pm .07$ Hz. Thus marginally within errors the shape of the
low frequency component does not seem to vary. For the high frequency
component of the low energy band the centroid frequency and width 
are $1.21\pm 0.3$ Hz and  $5.7\pm 0.2$ Hz, 
while for the high energy band they are  $1.8\pm 0.35$ Hz and  
$6.5\pm 0.35$ Hz.  Note that for a broad lorentzian the
peak of the frequency times power spectrum as shown in Figure
\ref{powspec} occurs not at the centroid frequency $f_c$ but 
at $(f_c + \sqrt{f_c^2 +(\sigma/2)^2})/2$ where $\sigma$ is the width. 
Thus, while the shape of the low frequency power spectral
component seems to be same with energy, 
there is evidence that the high frequency ones
changes and peaks at a higher frequency at higher energies. The
strength of the components (i.e. their r.m.s) changes with energy
especially their relative strength. This is shown more clearly in
Figure \ref{rmsene} where the r.m.s of the two components are plotted
as a function of energy. Here, the two component lorentzian model has
been fitted to the power spectra of each energy bin. Clearly, while
the r.m.s of the low frequency component decreases rapidly with energy
that of the high frequency one is relatively constant. We note that while there
maybe uncertainties on the absolute value of the r.m.s 
(due to uncertainties in the actual background level), the relative strength
of the two is independent of any background variation which will effect both
in the same manner.

\begin{figure}
\centering
\includegraphics[width=0.34\textwidth,angle=-90]{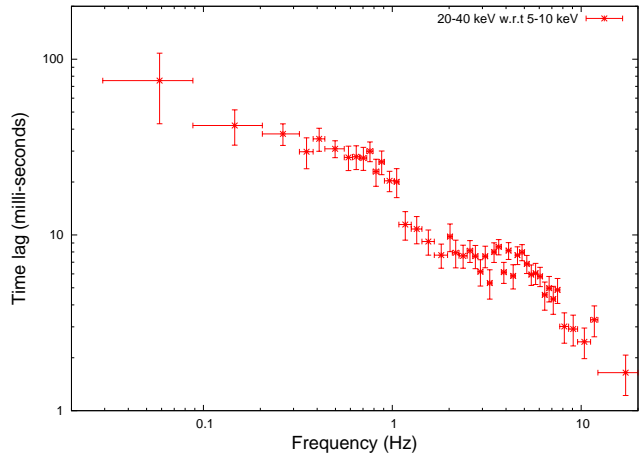}
\caption{The time lag of the 20-40 keV photons with respect to the
5-10 keV photons as a function of frequency. The behavior seems
to correspond to the power spectra shown in Figure \ref{powspec}  }
\label{lagfreq}
\end{figure}

\begin{figure}
\centering
\includegraphics[width=0.34\textwidth,angle=-90]{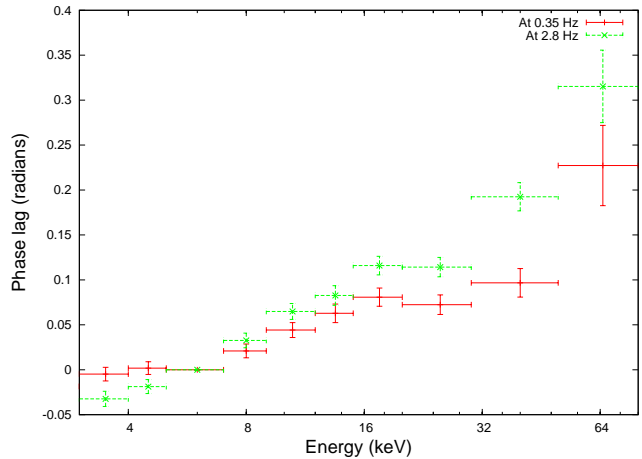}
\caption{The phase lag (i.e. $2\pi f$ times the time lag) 
of the photons in different energy band
with respect to those in 5-7 keV band for the low ($\sim 0.35$ Hz) and
for the high ($\sim 2.8$ Hz) frequencies.  }
\label{lagenergy}
\end{figure}

Figure \ref{lagfreq} shows the time lag of the high energy 20-40 keV photons
relative to the low energy 5-10 keV band as a function of frequency.
{  Just as in the case of the power spectra, the 
lightcurve was divided into 702 segments each having a 
length of 1024 points with a time bin of 16.67 milli-secs
and cross spectrum for each segment was computed and averaged. 
The time lags and their error were 
computed from the averaged cross-spectra using the scheme described 
in \cite{Now99}. The corresponding time-lags were then
rebinned in  frequency space.}
The time lag as a function of frequency for Cygnus X-1 has known to 
be roughly a power-law \citep{Now99}, 
but they were indications of complexities such as ``steps'' or
regions in frequency space where the time-lag is roughly constant. Here,
the LAXPC data provides {  confirmation}.
The time lag is nearly constant between $0.05$ to $\sim 0.9$ Hz at
$\sim 50$ milli-seconds and then drops to around $\sim 8$ milli-seconds 
in the frequency range $2$ to $6$ Hz before finally dropping to 
$\sim 3$ milliseconds at higher frequency. {  As indicated in
the RXTE analysis, 
the time lag behavior with frequency has a close correspondence with
the power spectra at the two energy bands, in the sense that the
behavior of the low and high frequency components are distinct.}

\begin{figure}
\centering
\includegraphics[width=0.34\textwidth,angle=-90]{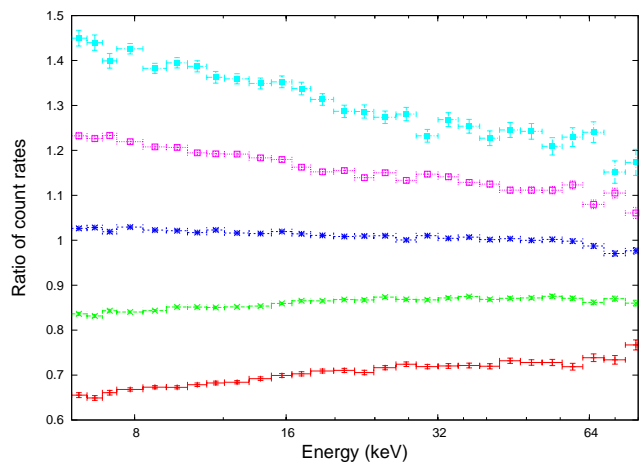}
\caption{The count rate ratio versus energy of  flux 
resolved spectra at 1 second time bins normalized to the average count rate.
The five spectra correspond to different fluxes corresponding to
 every alternate value shown in Figure \ref{indexflux}.
The spectra clearly show softening with increasing flux and with a pivot
point which is at high energies.  }
\label{countratio}
\end{figure}

\begin{figure}
\centering
\includegraphics[width=0.34\textwidth,angle=-90]{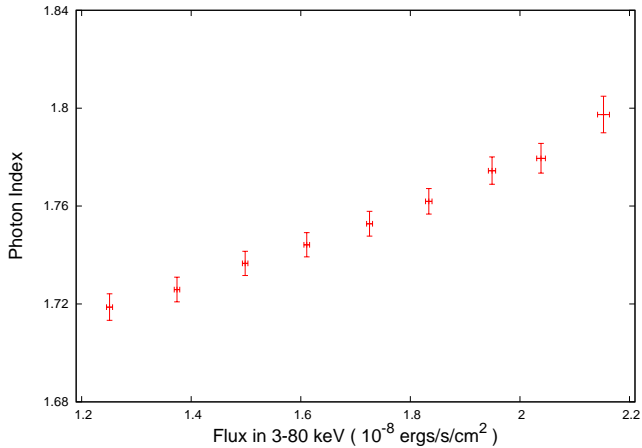}
\caption{The photon index versus flux for flux resolved spectra at a
time-scale of 1 second. }
\label{indexflux}
\end{figure}

Figure \ref{lagenergy} shows the phase lag at the peaks of
the low ($\sim 0.4$ Hz) and high ($\sim 3$ Hz) frequency components
as a function of energy. Here the reference is the 5-7 keV band.
The phase lags vary roughly linearly with the logarithm of the energy
as has been known before \citep{Now99}, except that the LAXPC data 
extends this relation to higher energies. It is interesting to
note that there seems to be some marked difference between the
low and high frequency behavior with the time-lag for low frequency component 
seems to show leveling off at high energies. However, the statistics
is not sufficient to make concrete statements.  

The event mode data of LAXPC allows flux resolved spectroscopy at
fast time-scales. The power spectra shown in Figure \ref{powspec}
shows that a natural time-scale which demarks the low frequency 
behavior from the high one, is about 1 second. The 
lightcurve binned at 1 second as  shown in Figure \ref{lightcurve}
can be divided into several flux (or count rate) values and spectroscopy
can be done on each of them.   Since, the spectral analysis will
be dominated by systematics (rather than Poisson statistics), we consider only
LAXPC 10 for the analysis.  Figure \ref{countratio} shows the
ratio of the count rates at five different flux levels divided
by the average count rates as a function of energy. Spectral
evolution as the flux changes is clearly visible and specifically
it can be seen that the source becomes softer as it gets brighter.
The pivot point where the variability should be zero, seems to
be at energies $> 80 $ keV. To quantify the spectral variation, we
fitted the different flux resolved  spectra with the same model used
for the average one, and plot the photon index versus the unabsorbed
3-80 keV {  unabsorbed} flux in Figure \ref{indexflux}. The photon index clearly shows
correlation with flux changing by nearly $0.1$ for a factor
of two variation in the flux, at a remarkably fast time-scale of 1 second.
The variation of the other spectral parameters such as those of the disk component
are within errors.  
Clearly tighter constraints can be obtained once the background and the
response of the detectors is better modeled.
{  The analysis shows that LAXPC data can be analyzed using more sophisticated
techniques like frequency resolved spectroscopy where spectra corresponding
to different frequency variation are generated \citep{Rev99,Rev00,Rei06} 
 and also by the alternate technique known as 
time scale resolved spectroscopy, where the analysis is undertaken in
the time domain \citep{Wu09}. Such an analysis can reveal spectral variations,
such as that the spectral index correlates with flux for Cygnus X-1,
on time-scales as short as 64 msecs \citep{Wu10}. However, these analysis 
have been limited to few of the RXTE observations taken in particular
modes and even for these data the spectral information was often
limited to few energy bins \citep{Wu09,Wu10}.}

\section{Discussion}

The event mode data of LAXPC having a significantly larger effective area at
energies above 30 keV than RXTE has provided an enhanced view of the
hard X-ray variability of Cygnus X-1 in its hard state. 

{  The timing behavior
can provide clues to the geometry of the system. For example,
frequency resolved spectroscopy of Cygnus X-1 in the hard state, suggests that
the disk is truncated \citep{Rev99}.  Here, we discuss the results, 
to indicate that they are 
consistent with the  qualitative features of 
the propagation fluctuation model in a geometry where the standard disk is
truncated and stochastic fluctuations occur
throughout  the disk (and the inner hot flow) at characteristic 
frequencies, which propagate inwards.  
However, it may be possible that the same broad qualititative timing features 
can also be produced in a geometry where the disk extends to the last stable
orbit as indicated by spectral fitting of the complex Iron line 
\citep{Par15}.  The timing results need to be
compared with detailed quantitative predictions of models with different
geometry to make clear statements. 
}

The power spectrum can be characterized as having 
two broad lorentzian features termed here as 
low ($\sim 0.4$ Hz) and high ($\sim 3.0$ Hz) frequency components. The r.m.s of the low frequency component decreases with energy. This is 
more clearly evident in the 1 second binned flux resolved spectra, which
shows that the spectra soften with increasing flux. The photon
index increased from   1.72 to 1.80 as the flux increased by a factor
of two. {  Such a steepening with flux is expected if the
variability is primarily due to increase in the soft photon flux.} This indicates that the low frequency variability is primarily 
driven by variation in the seed photon flux into the Comptonizing medium
which is likely to be the outer truncated standard disk.  These low frequency
fluctuations would then propagate inwards on viscous time-scale causing 
variations in the inner corona such as in its heating rate. Thus the low
frequency variability of the thermal Comptonization component has two   
drivers. First the variation of the seed photon flux (causing the
spectrum to soften with increasing flux) and later after the
fluctuation has reached the inner flow a variation in the overall flux
of the corona. Thus one expects a time delay between the two drivers
which may show up as the observed 
time-lag between the hard and soft X-ray photons.   

For the high frequency component, the r.m.s decreases less rapidly with energy
as compared to the soft one. Moreover there is some evidence that 
the shape of the component changes with energy in the sense that its peak 
shifts to a higher frequency as the energy increases. The time-lag between the
high and and low energy photons is significantly smaller than the case
for the low frequency component. All this seem to indicate that the
high frequency component is driven by variations in the thermal plasma
itself i.e. it originates in the inner flow in accordance with the
fluctuation propagation model. 

The results presented here showcase the unprecedented view that LAXPC
provides of the rapid variability properties of the X-rays above 30 keV
for Cygnus X-1. Clearly, further observations of the source in different
spectral states when combined with data from other instruments of
{\it AstroSat} will allow quantitative testing of models that strive to
explain variability properties of such systems and hence to constrain the
dynamics of the inner accretion flow near the black hole.           

\section{Acknowledgments}
We acknowledge the strong support from Indian Space Research
Organization (ISRO) in various aspect of instrument building, testing,
software development and mission operation during payload verification
phase. We  acknowledge support of TIFR central workshop during the
design and testing of the payload.

\end{document}